**ORIGINAL ARTICLE**

# High efficiency transmission grating for the ESO CUBES UV spectrograph

**Uwe D. Zeitner[1]** · **Hans Dekker[2]** · **Frank Burmeister[1]** · **Thomas Flügel-Paul[1]** · **Andrea Bianco[3]** · **Alessio Zanutta[3]**

**Abstract**
CUBES is the Cassegrain U-Band Efficient Spectrograph, a high-efficiency instrument operating in the UV spectral range between 300 nm and 400 nm with a resolution not less than 20,000. CUBES is to be installed at a Cassegrain focus of the Very Large Telescope of the European Southern Observatory. The paper briefly reviews various types of devices used as dispersing elements in astronomical spectrographs to achieve high resolution, before identifying binary transmission gratings produced by microlithography as the best candidate technology for the CUBES instrument. We describe the lithographic fabrication technology in general, two different design considerations to achieve the required high-resolution transmission grating, its prototyping by a direct-write lithographic fabrication technology, and the characterization of the achieved optical performance. An outlook to the realization of the grating for the final instrument, taking the most recent developments of lithographic writing capabilities into consideration is given.

## 1 Introduction

The Top-Level Requirements for CUBES [11] call for a spectral resolution R > 20,000 and "highest possible instrumental efficiency" in the UV. A comparison of the spectral resolution and efficiency of various types of dispersing elements for an 8-m telescope is made in [7], see Fig. 1.

✉ Uwe D. Zeitner
uwe.zeitner@iof.fraunhofer.de

1  Fraunhofer Institute for Applied Optics and Precision Engineering, Jena, Germany

2  Munich, Germany

3  INAF - Osservatorio Astronomico di Brera, Merate, Italy

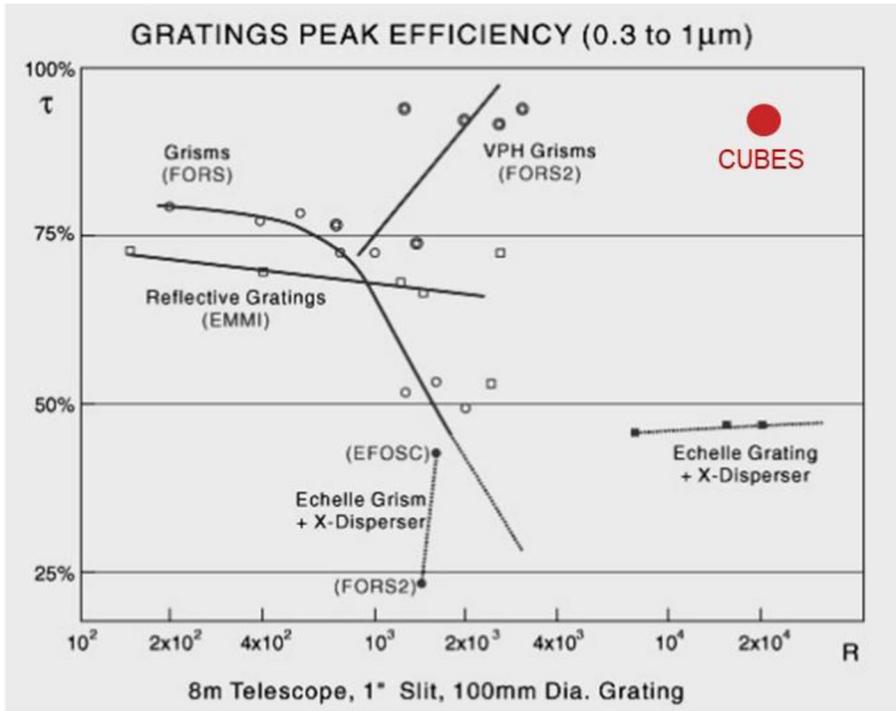

**Fig. 1** Peak optical efficiency of ESO gratings as a function of spectral resolution for an 8-m telescope, 1 arc sec slit and a 100 mm collimated beam, figure from [7]. The requirements for CUBES are indicated by the red circle

The requirements for CUBES imply a disperser system that is located in the top right-hand corner of the diagram.

Normally, astronomical high-resolution spectrographs are realized with a reflective Echellé grating that is combined with a cross-disperser to separate the orders on the detector. Considering the losses on the cross-disperser and the variation of diffraction efficiency along the orders, the typical average efficiency of this grating combination is 45% at best (see Fig. 1).

First-order reflection gratings do not deliver the desired diffraction efficiency.

and in addition are difficult to incorporate in a compact optical design that is suitable for a Cassegrain focus.

On the other hand, first-order Volume Phase Holographic Gratings (VPHG, see e.g. [2]) are efficient and allow a compact optical design since incident and diffracted beams need not be separated. However, VPHG are not efficient in the UV below ~350 nm due to absorption and scattering in the dichromated gelatine layer.

In search for an alternative technology to produce transmission gratings, the lithographic grating technology used for manufacturing the Gaia grating appeared promising to us [12].

**Table 1** Specification of the prototype grating

| Type of grating | binary transmission grating |
| --- | --- |
| Line density | 3448 lines/mm |
| Grating size | 250×130×15 mm |
| Operating wavelength range | 300–400 nm |
| Angle of incidence | 31.27° |
| Diffraction efficiency | best effort, blazed for 320 nm |
| Polarization independency | best effort |
| Wavefront quality and substrate wedge | not specified |
| A/R coating on non-grating side | not required |

Studies on CUBES commenced as early as 2012 [1] and in 2014 funding to manufacture a prototype grating using this technology was obtained by Prof. B. Barbuy (IAG, Sao Paolo University) from FAPESP, the São Paulo Research Foundation. Its requirements were based on an early 3-slicer strawman design of CUBES with a slicer slit width of 0.375″, a collimated beam size of 220 mm and a beam footprint on the grating of 220 mm×250 mm. This design covered the spectral range 300 nm – 400 nm with good efficiency in the UV part of the spectrum. The large beam size implied a large and heavy instrument that would likely severely strain the load-carrying capability of a VLT Unit Telescope.[1]

Because the required grating size exceeded the manufacturing capabilities of IOF, the design foresaw a mosaic of 250×260 mm consisting of two transmission gratings, mounted in a common frame. The motivation for the prototype development was proving that with this technology high UV diffraction efficiency could be achieved over the whole area of one segment. The requirements of the segment prototype are listed in Table 1.

The contract was placed mid-2016, the grating was completed 1 year later. It was fully characterized by the end of 2018.

In this paper we will describe the fabrication technology, optical design and manufacturing of the prototype, its optical characterization, and the development plan for the next phases of CUBES.

## 2 Lithographic grating fabrication technology

Historically, the fabrication of diffraction gratings for spectroscopic instruments was mainly based on either mechanical ruling or holographic recording of a two-beam interference pattern in a photosensitive polymer [5]. Mechanical ruling of gratings can achieve very high line densities (grating grooves per millimeter) but takes a very long processing time due to the sequential process [6]. Consequently, the costly ruled grating structures are typically not used directly but as a master in a subsequent replication step transferring the grating profile into a polymer layer on a rigid substrate.

---

[1] The requirements for Cassegrain instruments call for an instrument mass not exceeding 2500 kg and a moment of inertia less than 20,000 Nm. For this reason, other Cassegrain instruments like FORS and the 3-arm XSHOOTER have beam sizes not exceeding 100 mm.

The holographic recording of the grating into a polymer is done in parallel for the full grating area but requires a precise control of the interfering wave-fronts in order to achieve a recorded grating pattern with low wave-front error. The resist mask can then be transferred into the underlying substrate by etching and gratings with high diffraction efficiency for at least one polarization can be achieved [8]. A very cost-effective version to obtain gratings with high diffraction efficiencies avoiding the transfer step is the holographic recording of the interference pattern in form of a refractive index change in a thin layer of a dedicated photo-sensitive material [10]. However, as already mentioned in Sec. 1 these VPHGs are not suitable for UV applications as needed for CUBES.

The Fraunhofer Institute for Applied Optics and Precision Engineering (Fraunhofer IOF) is developing various types of high-resolution gratings using a different technology which is the generation of the grating pattern by a sequential direct write lithographic exposure of a resist and a subsequent transfer of the resist mask into an underlying substrate or layer stack [3]. This lithographic fabrication process is derived from the integrated circuit manufacturing in microelectronics and adapted to optical applications. A very typical substrate material used for micro-optical components is fused silica or $SiO_2$, a material very well suited also for UV-applications.

The direct write lithographic fabrication approach, however, has not been used for the realization of high-performance spectroscopic gratings in the past. The reason is related to the sequential writing regime of the lithographic exposure [13]. To fill a larger grating area, many laterally limited writing fields need to be stitched to each other with extremely high precision. Otherwise, small positioning tolerances in the range of 1/20…1/50 of the grating period or even below will result in severe disturbances of the grating function and the occurrence of so called "grating ghosts". These are spurious diffraction peaks that might be mistaken for real spectroscopic lines e.g. in an emission spectrum. In 2006 a novel electron-beam lithography tool of type SB350 OS (Vistec Electron Beam GmbH) was installed at the Fraunhofer IOF. In addition to its high patterning resolution below ~65 nm feature size and its positioning accuracy of ~13 nm (3σ rms), an important characteristic is the variable-shaped-beam (VSB) writing principle, which allows use of the high-resolution patterning capability on large areas. The largest substrate dimension that can be accommodated in the tool is 292 mm × 150 mm × 15 mm. Its suitability for the realization of high-performance spectroscopy gratings has been proven during the development of the Radial-Velocity-Grating for the *Gaia* satellite of the European Space Agency [12]. This blazed grating covering an area of 155 mm × 205 mm was realized as a binary effective-refractive-index sub-wavelength structure etched into a fused silica substrate. The high lateral resolution and accuracy of the electron-beam writing process was mandatory to achieve the required efficiency-, straylight-, and wave-front performance.

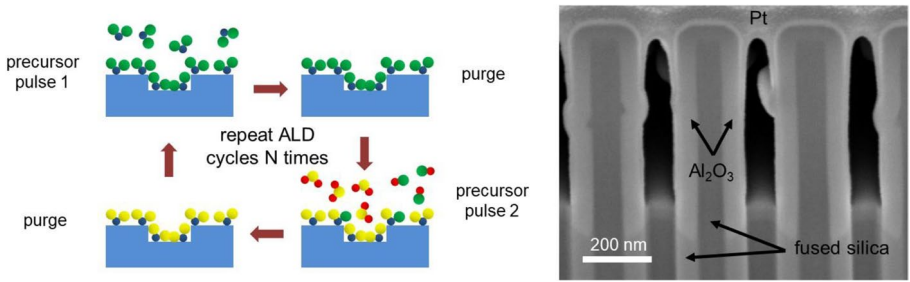

**Fig. 2** Left: Schematic of the atomic-layer-deposition (ALD) process. Right: It is possible to realize conformal overcoatings with high refractive index materials (e.g. $TiO_2$, $Al_2O_3$, $HfO_2$, …) of micro-structures already etched into a substrate. The Pt-layer on top of the grating structure was only applied for the preparation of the cross-section image

A major challenge towards the use of high dispersive surface relief gratings for spectroscopic applications is the need for a high, polarization independent diffraction efficiency. The dispersion of a grating can be directly derived from the grating equation to be

$$\frac{\partial \alpha_m}{\partial \lambda} = \frac{m}{n \bullet p \bullet \cos \alpha_m(\lambda)},$$

with $\lambda$ being the wavelength, $p$ the grating period, $n$ the refractive index in the space behind the grating, and $\alpha_m$ the diffraction angle of the $m^{\text{th}}$ diffraction order. According to this equation, high dispersions can only be achieved for a large $m$ or small grating periods in the range of the wavelength or even below. The use of large diffraction orders $m$ is utilized in Échelle-gratings but require the use of a cross-disperser to separate the spectra of different orders from each other. In contrast, the use of a grating in 1st diffraction order at a period $p \sim \lambda$ physically suppresses higher diffraction orders and thus, forcing the radiation to only the 0th or 1st diffraction order which enables high diffraction efficiencies. However, gratings in this so-called *resonance domain* typically exhibit a very strong polarization dependency of the diffraction efficiency [4]. To overcome this difficulty additional parameters for the functionalization of the structure within the grating period are needed. Without going too much into the details of rigorous grating design considerations as this is beyond the scope of this paper, there are two options to tackle the polarization problem. The first one is to etch the grating structure into a layer having a high refractive index, resulting in a comparable shallow grating profile. Alternatively, it is possible to add a high-refractive index layer on top of the comparably low refractive index $SiO_2$ grating bars. A suitable technology for such an overcoating is atomic-layer-deposition (ALD). It is a specific version of a chemical vapor deposition (CVD) process which consists of a sequence of repeated, surface activated coating and purging steps. A schematic of the ALD process is shown in Fig. 2. Due to its coating characteristic, it forms a conformal layer of almost equal thickness on top of the already present surface topography (see right side of Fig. 2. Note: the arrows pointing to fused silica material are correct. The image contrast is not determined by the material but by the so-called "curtaining"-effect caused by the FIB-cut preparation).

This approach opened the way to drive the direct write lithographic fabrication of SiO$_2$ gratings towards shorter wavelengths and design gratings exhibiting a high dispersion while simultaneously achieving a high *and* polarization independent efficiency [14]. This was an important pre-requisite for the development of the CUBES prototype grating.

## 3 Optical design and manufacturing of the grating prototype

The specifications of the full-size grating are summarized in Table 1. To realize the best optical performance especially with respect to the achievable diffraction efficiency within the specified spectral range, two different grating configurations were developed during the design phase of the project.

The first design is based on an early version of the gratings fabricated as small-size prototypes already before the project for a full-size proof-of-concept grating was started. It relies on the conformal coating of medium refractive index layer of Al$_2$O$_3$ by atomic layer deposition (ALD) on the grating structures etched into the fused silica substrate having a lower refractive index. This material combination was chosen to reduce the polarization sensitivity of the diffraction efficiency compared to a pure fused-silica structure.

Since a discrepancy between theoretical and measured diffraction efficiencies occurred for the small-size prototypes, an alternative design was investigated, comprising a binary grating structure etched directly into a high refractive index layer of silicon nitride (Si$_3$N$_4$). Both designs are discussed in the following.

### 3.1 Grating design considerations

#### 3.1.1 Design 1: Binary grating in fused silica and ALD of Al$_2$O$_3$

The grating configuration of Design 1 is illustrated in Fig. 3. The binary grating structure is etched into the fused silica substrate with a depth of 637 nm and a groove width of 225 nm. After lithographic fabrication, the grating profile is coated with a conformal layer of Al$_2$O$_3$ of 56 nm thickness. The conformal layer characteristic is achieved by using ALD for coating of Al$_2$O$_3$.

Figure 4 shows the optical performance of Design 1 within the operating spectral range. The design and theoretical performance evaluation has been performed with an in-house developed rigorous-coupled-wave analysis (RCWA) software. The polarization averaged efficiency $\eta_{av}$ as well as the polarization sensitivity $PS$ are determined by the efficiencies $\eta_{TE}$ and $\eta_{TM}$ for TE- and TM polarization according to

$$\eta_{av} = \left(\eta_{TE} + \eta_{TM}\right)/2$$

and

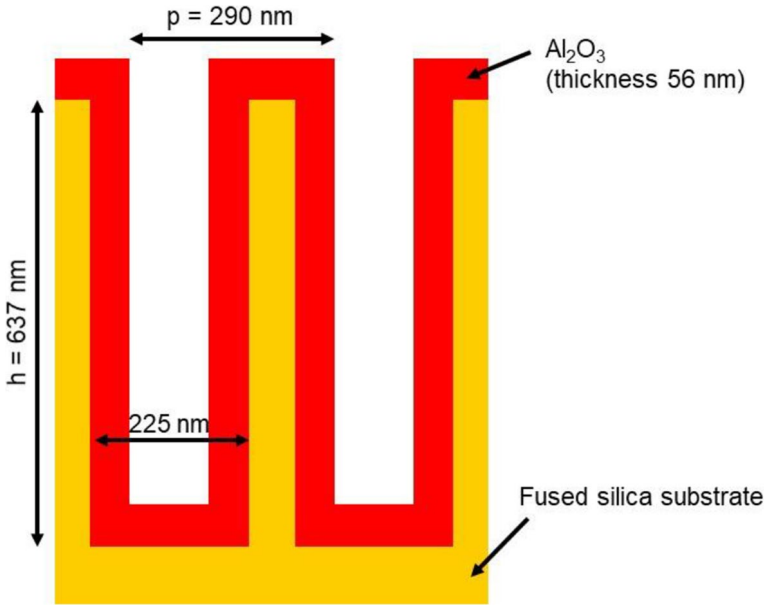

**Fig. 3** Grating geometry according to Design 1. The binary grating structure is etched into a fused silica substrate and conformally coated with $Al_2O_3$

$$PS = \frac{\eta_{TE} - \eta_{TM}}{\eta_{TE} + \eta_{TM}}.$$

From Fig. 4 it can be seen that $\eta_{av}$ is well above 85% from 300 nm to 350 nm and decreases to 60% at 400 nm wavelength. The polarization sensitivity is less or equal 6% with a maximum value of 6% at 400 nm.

### 3.1.2 Design 2: Binary grating in $Si_3N_4$

Design 2 is based on a binary grating structure etched into a $Si_3N_4$ layer with a thickness of 242 nm. The groove width of the optimized grating structure is 173 nm. The corresponding grating geometry is sketched in Fig. 5. For lithographic fabrication of this grating configuration, first, a layer of $Al_2O_3$ with a thickness of 15 nm needs to be deposited on the fused silica substrate followed by a layer of $Si_3N_4$ layer with a thickness of 242 nm on top. The $Al_2O_3$ layer is used as a stop layer for etching of the grating structure into $Si_3N_4$ in order to prevent etching into the underlying substrate.

The optical performance of Design 2 is summarized in Fig. 6. The polarization-averaged efficiency is above 78% over the complete operating spectral range with minimum values of 79% at 300 nm and 78.3% at 400 nm. The corresponding polarization sensitivity is below 8% over the full spectral band with maximum values of 5% at 300 nm and 7.3% at 400 nm.

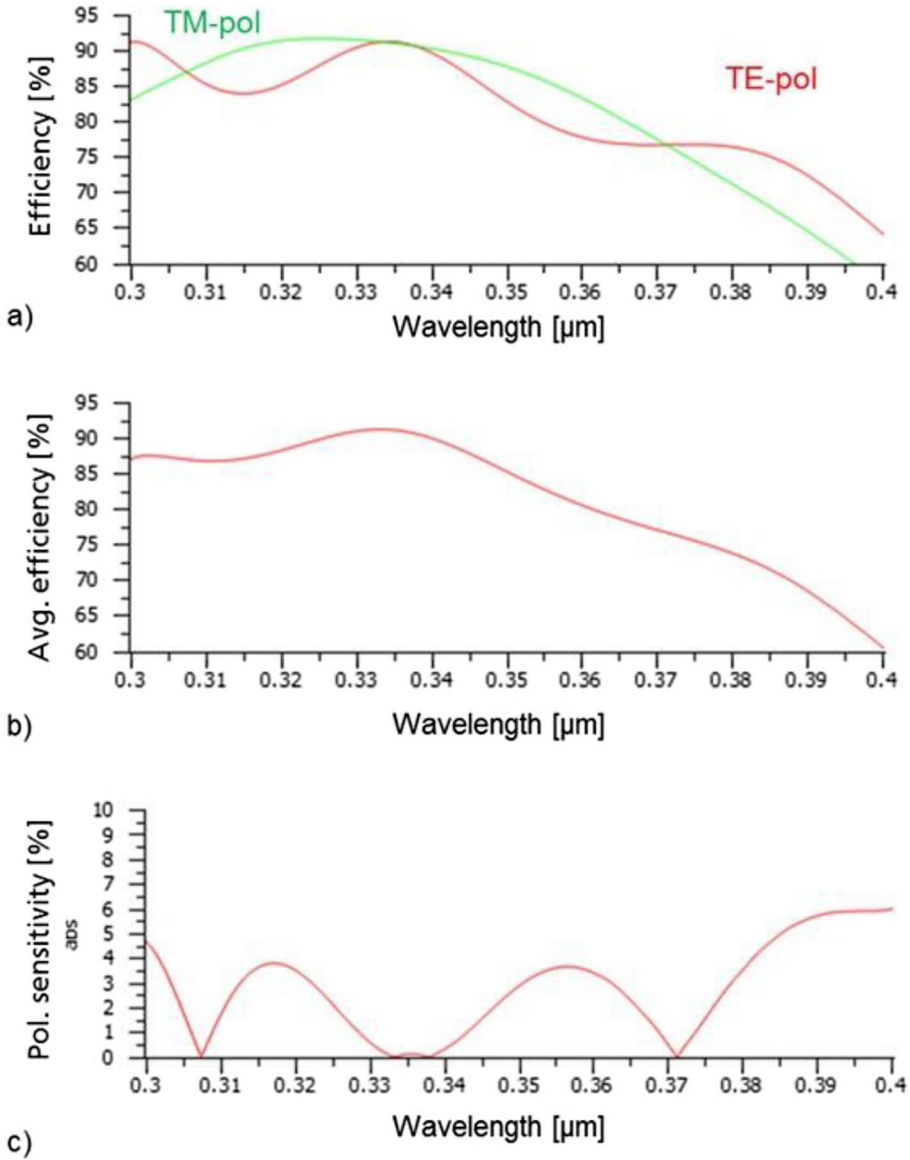

**Fig. 4** Optical performance of Design 1 within the operational spectral range. **a**) diffraction efficiency for TE- and TM-polarization, **b**) polarization averaged diffraction efficiency, **c**) polarization sensitivity

By comparing the optical performance of both designs it can be seen that the configuration etched into the $Si_3N_4$-layer shows a slightly higher average diffraction efficiency, especially for wavelengths above 340 nm. However, at wavelengths near the short wavelength end of the spectrum the efficiency of Design 2 drops more strongly than that of Design 1. This is caused by the inherent material absorption of $Si_3N_4$ in the UV. Furthermore, the evaluation of test coatings of $Si_3N_4$ deposited by low-pressure chemical vapor deposition (LPCVD) has revealed an even stronger absorption of up to 30% for the nominal layer thickness for wavelengths below 400 nm. This would substantially degrade the efficiency performance of the grating. Therefore, Design 2 was not considered applicable for the realization of a full-size demonstrator without a further detailed evaluation of other $Si_3N_4$ deposition methods and suppliers potentially being able to provide layers with lower absorption and sufficient homogeneity over the full grating area.

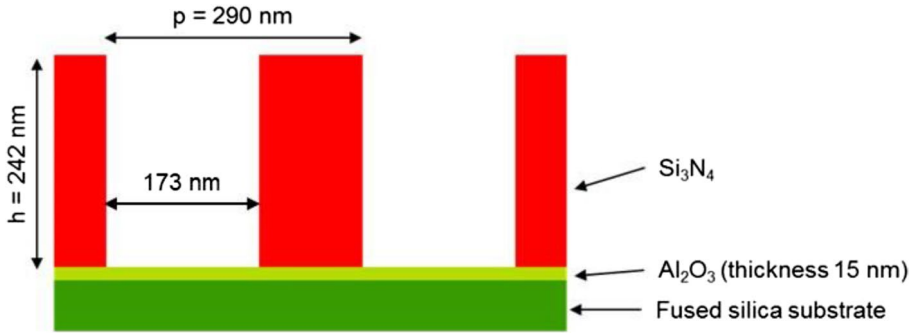

**Fig. 5** Grating geometry according to Design 2. The binary grating structure is etched into $Si_3N_4$. The $Al_2O_3$ layer on top of the substrate is used as etch stop for etching of the grating structure into the $Si_3N_4$ layer

### 3.2 Grating fabrication

The lithographic process chain for fabrication of the full-size grating is sketched in Fig. 7. Fabrication was based on electron-beam lithography followed by a reactive ion-etching process (RIE) to transfer the structure into fused silica. In between the individual processes, several characterization steps of the grating structure in its current state are performed. Most importantly, measurement of the duty-cycle with the use of a scanning electron microscope (SEM) and measurement of the final etching depth by use of atomic force microscopy (AFM) were done.

The groove width of the final grating is determined by the critical dimension (CD) of the chromium mask, which covers the grating bars during deep etching. Therefore, control of the CD of the chromium mask before deep etching is crucial for fabrication of a grating complying with the design. Figure 8 shows a SEM picture of the chromium mask in the center of the substrate. The measured CD of the mask openings was 229 nm, which is in very good agreement with the nominal value of 225 nm. Due to the large size of the substrate, SEM pictures can only be taken in a central position on the substrate. However, characterization of the chromium mask in the outer regions can also be done indirectly, by measuring the optical transmission in the 0th diffraction order of the Cr-amplitude grating. The measured transmission depends on the thickness of the chromium mask and the local CD. The thickness of the chromium mask was measured after removal of the resist by using an AFM with a result of $(41 \pm 2)$ nm over the complete grating area and was consequently assumed to be constant.

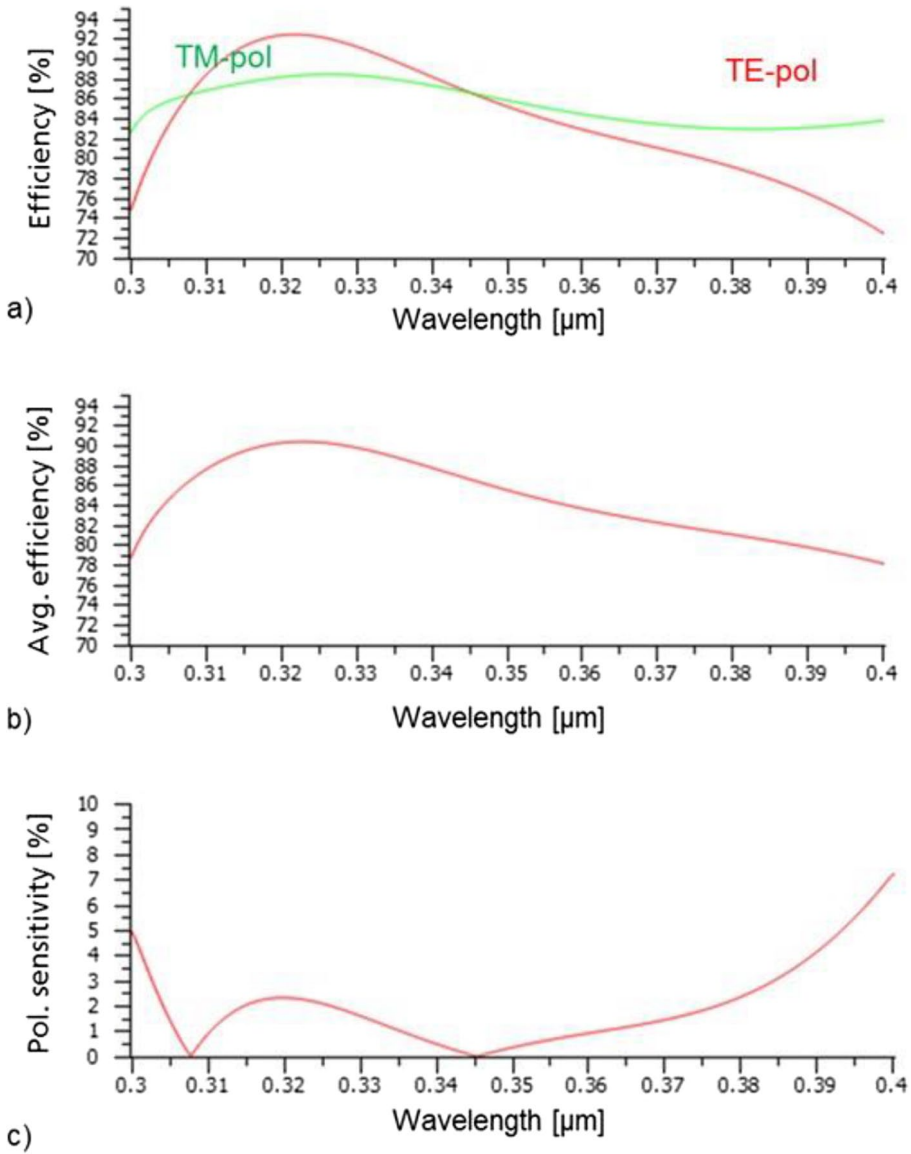

**Fig. 6** Optical performance of Design 2 within the operational spectral range. **a**) diffraction efficiency for TE- and TM-polarization, **b**) polarization averaged diffraction efficiency, **c**) polarization sensitivity

Measurement of the transmission in the 0th diffraction order was performed at a wavelength of 375 nm and an angle of incidence (AOI) of 20° for TE and TM-polarization, respectively. The AOI of 20° was chosen to achieve a high sensitivity

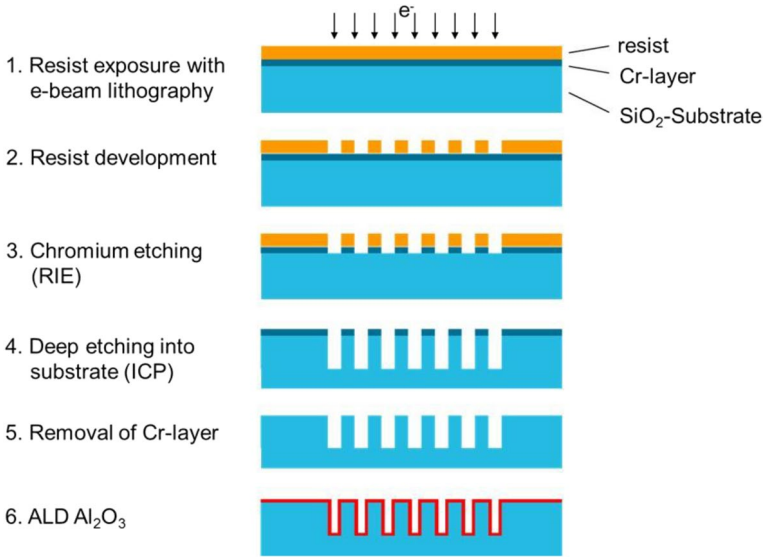

**Fig. 7** Schematic of the lithographic process chain for the fabrication of the full-size grating

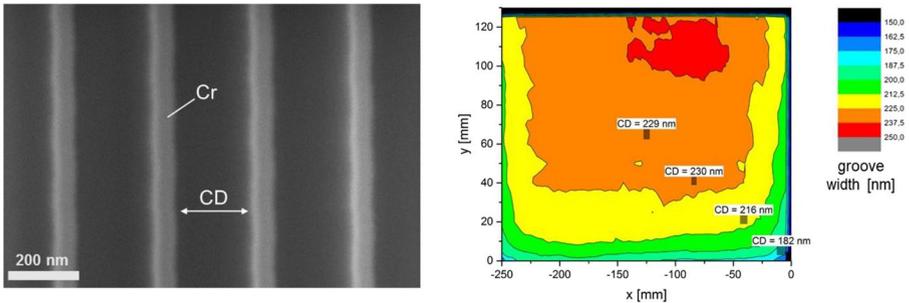

**Fig. 8** Left: SEM picture of the Cr-mask in a central position on the substrate. Right: CD of the chromium mask over the grating area. Values were calculated from measurement of the optical transmission in the 0th diffraction order. Please note the different scaling of x- and y-axis

of the transmission dependence of the CD. The grating area was scanned in x- and y-direction in 5 mm steps. The CD of the chromium mask was calculated from the measured transmission values using rigorous coupled-wave analysis (RCWA). The right side of Fig. 8 shows the resulting contour map of the CD over the grating area. The CD varies from 229 nm in the center of the grating to 189 nm at the marginal area. This variation is due to the large size of the grating and the associated inhomogeneities of individual process steps, e.g. the resist baking after its application.

After fused silica etching and removal of the chromium mask, the etching depth over the grating area was measured with an AFM (Nanostation 500). Since the variation of the etching depth shows a radially symmetric characteristics due to the geometry of the etching tool's plasma chamber, the measurement was performed on four positions along a diagonal over the grating area beginning from the center. Figure 9 shows distribution of the measurement positions over the grating area and the corresponding depth values.

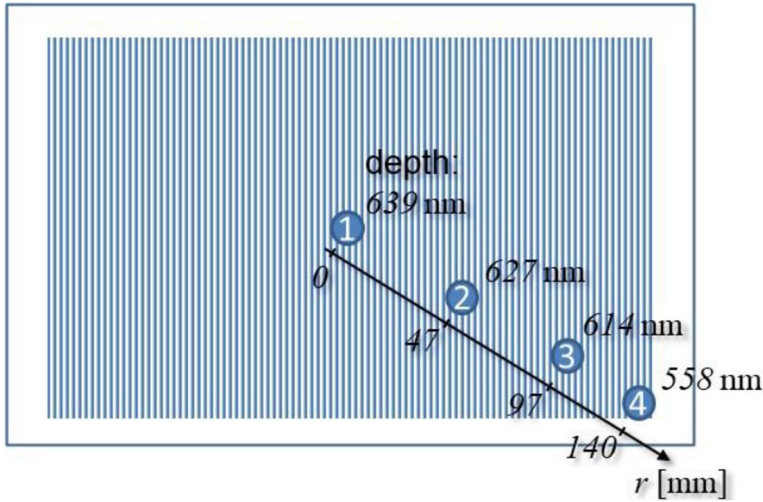

**Fig. 9** Positions and etching depth values measured by AFM

**Table 2** Values of the re-optimized $Al_2O_3$ layer thickness at different positions over the grating area

| Position | Radius [mm] | CD [nm] | Etching depth [nm] | Reoptimized $Al_2O_3$ thickness [nm] |
|---|---|---|---|---|
| 1 | 0 | 229 | 639 | 58 |
| 2 | 47 | 230 | 627 | 58 |
| 3 | 97 | 216 | 614 | 46 |
| 4 | 140 | 182 | 558 | 26 |

Based on the results from structural characterization, the layer thickness of $Al_2O_3$ was re-optimized with respect to highest diffraction efficiency and uniformity over the grating area. In a first step, the $Al_2O_3$ thickness was optimized individually for each of the four positions shown in Fig. 9, based on the locally measured CD and etching depth. The results are summarized in Table 2. The values of the optimized $Al_2O_3$ layer thickness vary from 58 nm in the center to 26 nm at the marginal region of the grating area. From the lateral distribution of the inhomogeneities it was derived that the best average performance over the full grating area can be achieved by applying an $Al_2O_3$ layer thickness of 50 nm. The resulting optical performance in the central region of the grating will be slightly decreased compared to the preliminary results, but performance in the

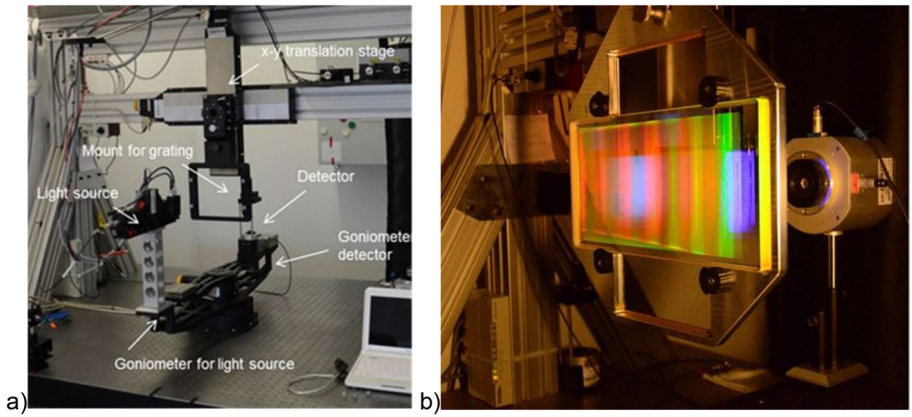

**Fig. 10** **a** Customized set-up for diffraction efficiency characterization of the gratings. **b** Full-size grating mounted in the setup for measurement of the diffraction efficiency

outer region will be improved strongly compared to the nominal layer thickness of 56 nm.

ALD of $Al_2O_3$ was carried out by the company Beneq (Finland). The ALD run was monitored with silicon witness samples placed around the substrate. The thickness determined from the witness samples was $(50 \pm 1)$ nm with a very good uniformity of below 0.5% over the full grating area. A scanning electron microscopy image of a grating cross section taken at a test grating is depicted in Fig. 2.

## 4 Characterization results

### 4.1 Optical characterization of diffraction efficiency homogeneity

The characterization of the polarization dependent diffraction efficiency of the CUBES grating prototype was performed using a customized diffraction efficiency measurement setup available at the IOF. A picture of the setup is shown in Fig. 10.

For a given incidence angle of the light source, the grating area is scanned with a collimated laser beam of about 2 mm diameter and the local power in a particular diffraction order is recorded using a Si-photodetector. The relation to the incident power then gives a map of the local diffraction efficiencies. In this setup the power of the detector is measured using an integrating sphere (Ulbricht sphere). Both power values, the one of the light source and one of the detector, are measured via a lock-in amplifier. The grating area was scanned in x- and y-direction in 5 mm steps which gives a detailed map of the local efficiencies over the entire grating area. The diffraction efficiency of the -1st order was measured separately for the two linear polarizations parallel and perpendicular to the grating lines (TE- and TM-polarizations, respectively). The polarization averaged efficiency $\eta_{av}$ as well as the polarization sensitivity $PS$ were calculated from the measured efficiencies $\eta_{TE}$ and $\eta_{TM}$ for TE- and TM polarization, respectively.

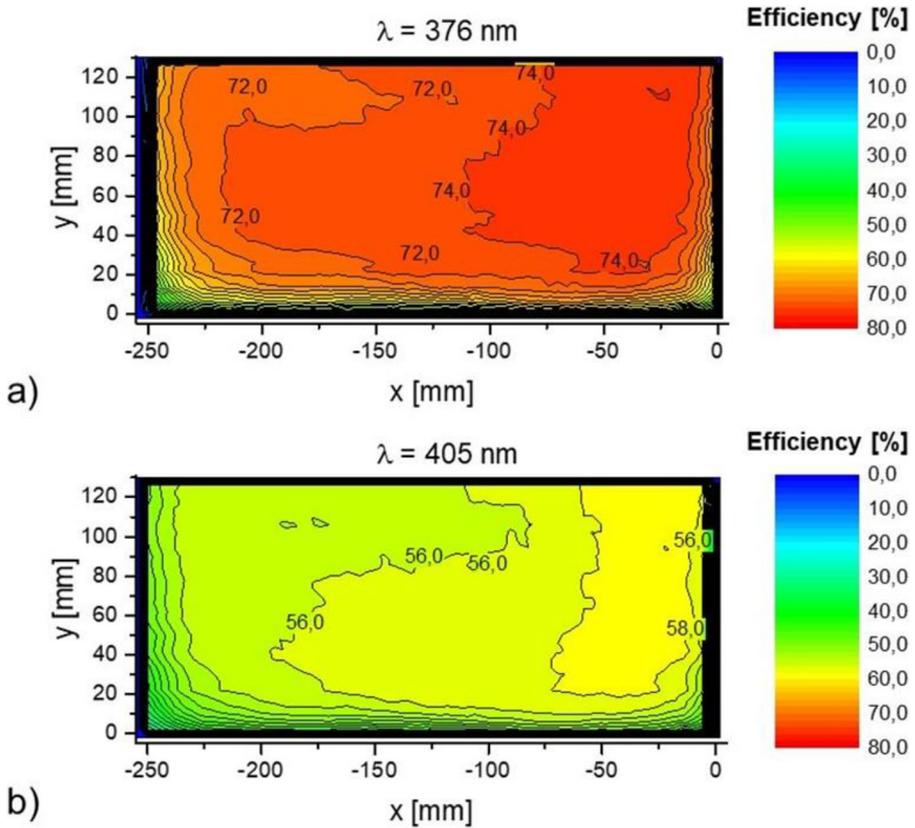

**Fig. 11** Measured polarization-averaged diffraction efficiency over the full grating area for wavelength of **a** 376 nm and **b** 405 nm

As light sources within or close to the operating spectral range, two laser diodes were available with wavelengths of 376 nm and 405 nm.

Figure 11 shows the measured polarization-averaged diffraction efficiency over the full grating area for 376 nm and 405 nm. Considering the large size of the grating, the measured efficiency shows a very good uniformity over the majority of the full grating area. The efficiency decreases only in the outer regions of the grating due to the decreasing CD and etching depth, which is manly caused by inhomogeneities of individual process steps during fabrication.

### 4.2 Diffraction efficiency over the spectral band

In addition to these surface-mapping measurements at the mentioned wavelengths, the diffraction efficiency at the center of the grating was measured at 11 discrete wavelengths between 300 and 410 nm at 10 nm intervals by PTB, the national metrology institute of Germany [9].

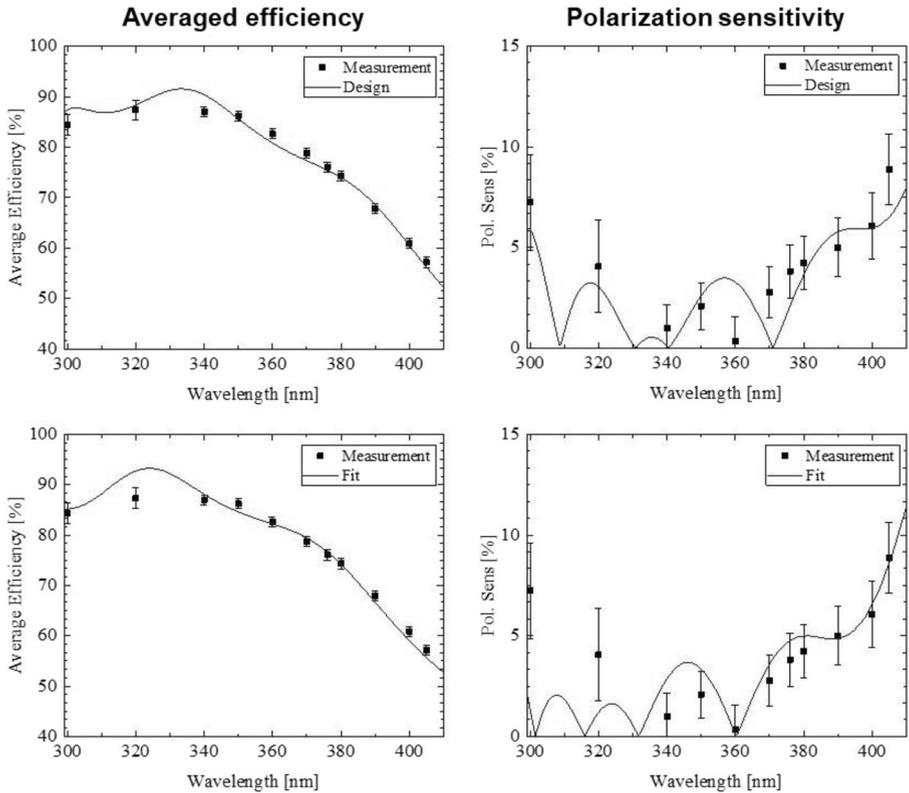

**Fig. 12** Wavelength dependent measured polarization-averaged diffraction efficiency and polarization sensitivity at central position of the grating. The upper and lower row of the figure shows the comparison of the measurement results with the corresponding curves derived from the initial grating design and the result of the fitting attempt, respectively

**Table 3** Nominal and fitted grating parameters. Fitted values correspond to the measured values of the diffraction efficiency

|  | CD [nm] | Etching depth [nm] | $Al_2O_3$ layer thickness [nm] |
|---|---|---|---|
| Nominal value | 225 | 637 | 50 |
| Fitted value | 224 | 620 | 52 |

The measurement points were used to fit the real grating parameters, namely CD, etching depth, and $Al_2O_3$ thickness. The spectral diffraction efficiencies of the fitted grating configuration were then calculated applying RCWA.

Figure 12 shows the measured values of the polarization averaged diffraction efficiencies and the related polarization sensitivity at the grating center. The curve fitted to the measured values is compared to the nominal design curve.

The fitted grating parameters as well as the nominal values are summarized in Table 3. As can be seen, measured and fitted optical performance of the full-size grating are both very close to the values of the nominal design. The fitted geometry shows a slightly better correspondence to the measurement at the red end of the spectrum while the nominal design corresponds slightly better to the measured values at the blue end of the spectrum. From that it can be concluded that the actual three-dimensional shape of the grating bars which is not considered in the calculations also has a slight influence on the actual grating performance.

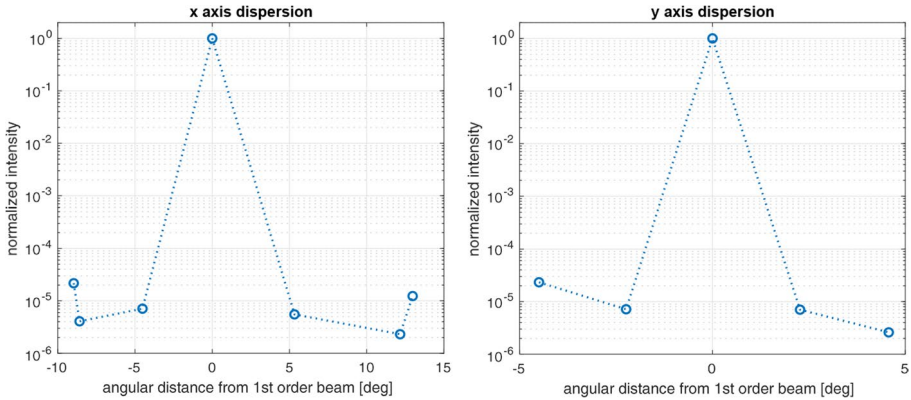

**Fig. 13** Relative intensity and angular location of grating ghosts in the dispersion direction (left) and in the perpendicular direction (right)

### 4.3 Stray light

Stray light measurements were carried out at INAF – Osservatorio Astronomico di Brera, using an S-polarized DPSS laser at 457 nm (Cobolt Twist, 200 mW).

The angle of incidence and diffraction were set up to be equal at 52° (Littrow condition). When placing a screen in the diffracted beam, one sees the bright first diffracted order as well as a line of "grass" and several faint grating ghosts in the dispersion direction, the first ones at about ±4.5° from the main peak. However, one also sees ghosts in the vertical ("cross") direction, the first ones at about 2.2° from the main peak. We have measured the relative intensity of the ghosts using a photodiode. The results are shown in Fig. 13. All ghosts are fainter than $2.10^{-5}$.

The "cross" ghosts with their angular distance resemble the pattern of diffraction orders of a weak and very coarse grating with an equivalent groove frequency of 53 l/mm (period 18.9 μm). This value is compatible with specific stitching parameters in the sequential E-beam writing process. There are strategies known to minimize this effect.

The intensity and angular distance of all ghosts, relative to the first diffracted order, are such that they are not expected to affect the quality of the science and calibration spectra in the CUBES spectrograph.

**Table 4** Preliminary specifications of CUBES gratings

| Grating | Blank size (mm) | Material | AOI (deg) | Wavelength range (nm) | Blaze wavelength (nm) | Line density (l/mm) |
|---------|-----------------|----------|-----------|-----------------------|------------------------|---------------------|
| #1 | 180×220×9 | Fused Silica | 36.07 | 302–352 | 320 | 3597.1 |
| #2 | | | 35.89 | 346–405 | 365 | 3115.3 |

## 4.4 Conclusion on test results

The test results on the prototype grating (that was optimized for 320 nm wavelength) show excellent control of groove shape and good efficiency in the range 300–350 nm. Homogeneity is good apart for some edge effects. As predicted, the average efficiency drops in the range 350–400 nm and the polarization increases to ~10% at 400 nm despite use of the $Al_2O_3$ ALD layer. Spectral ghosts are well below the $10^{-4}$ level that is generally considered acceptable for mechanically ruled gratings. An unexpected finding is the presence of faint ghosts perpendicular to the main dispersion axis.

## 5 Development plan for the next phases of CUBES

The optical design that was adopted as result of the Phase A instrument study employs two gratings. Compared to the one-arm design for which the prototype was made, advantages of the two-arm design with regard to grating performance are much higher optical efficiency, especially at the red end, and lower polarization. The gain in overall grating efficiency more than compensates the light loss in the dichroic beam splitter required for the two-arm design.

There are other gains at the spectrograph design level as well: lower variation in dispersion, sampling and spectral resolution due the reduced spectral coverage in each arm, the possibility to use two 9 k CCDs in standard cryostats instead of an 18 k or 24 k CCD mosaic that is mounted in a non-standard cryostat, and avoidance of spectral gaps between CCD array elements.

The beam and grating size were also reduced to keep instrument weight under control and to avoid the use of a grating mosaic. More details can be found in [11].

The preliminary grating parameters of the new design are summarized in Table 4.

We have performed simulations based on these parameters. The predicted diffraction efficiency is reported in Fig. 14 for both gratings with two manufacturing options: pure Fused Silica ($SiO_2$) grooves or with an $Al_2O_3$ ALD overcoating.

We hope to keep losses due to reflection on the non-grating surface and grating manufacturing errors (groove shape, homogeneity over the surface) at the level of 5%. In this case the average actual efficiency of the gratings in each band will be around 88%.

Grating #1:

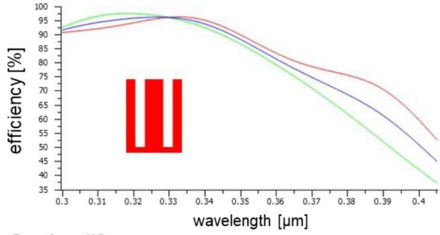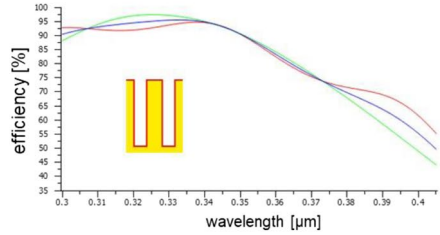

Grating #2:

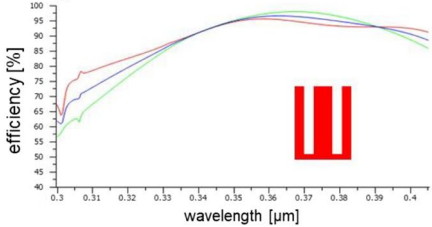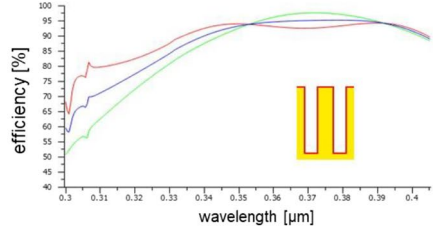

**Fig. 14** Theoretical diffraction efficiency of grating #1 (upper rows) and #2 (lower rows) with the two manufacturing options: solid Fused Silica and 15 nm ALD Al$_2$O$_3$ coating. The three curves represent TE (red), TM (green), and average (blue) efficiency

Since the groove width of grating #1 is very close to what is thought feasible based on experience with the prototype grating, manufacturing of a full-size prototype grating #1 is planned for 2022. It will be tested for stray light, diffraction efficiency and -homogeneity, and wavefront quality.

This prototype will be manufactured using the current facility at Fraunhofer IOF which can accommodate blanks of $230 \times 230 \times 9$ mm dimension. The blank can be cut to 180 mm after processing of the grating structure. IOF are currently planning an upgrade of its electron beam lithography tool to be installed towards the end of 2023. With this upgrade the maximum substrate dimension that can be accommodated in the machine will increase to about $300 \times 275 \times 15$ mm. The increased thickness would help to alleviate concerns about blank deformation and wavefront errors due to mounting stress and changing gravity.

The technology upgrade will also improve the resolution and positioning of the writing process by about a factor 2 and thus, will contribute to even lower stray light and wavefront error levels of the gratings. The intensities of the stray light ghosts in cross dispersion direction are expected to be reduced, too.

**Funding** Open Access funding enabled and organized by Projekt DEAL. Grating manufacturing was funded by the São Paulo Research Foundation (FAPESP), grant number 2014/18100–4.

**Data availability** Data underlying the results presented in this paper are not publicly available at this time but may be obtained from the authors upon reasonable request.

## Declarations

**Conflict of interest** The authors declare no conflicts of interest.

## References

1. Barbuy, B., Bawden Macanhan V., Bristow, P. et al. :CUBES: Cassegrain U-Band Brasil-ESO Spectrograph. Astrophys. Space Sci. **354**, 191 (2014)
2. Barden, S.C. , Arns, J.A., Colburn, W.S.: Volume-phase holographic gratings and their potential for astronomical applications. Proc. SPIE **3355**, 866-678 (1998)
3. Clausnitzer, T., Limpert, J., Zöllner, K., Zellmer, H., Fuchs, H.-J., Kley, E.-B., Tünnermann, A., Jupé, M., Ristau, D.: Highly efficient transmission gratings in fused silica for chirped-pulse amplification systems. Appl. Opt. **42**, 6934-6938 (2003)
4. Clausnitzer, T., Kämpfe, T., Kley, E.-B., Tünnermann, A., Tishchenko, A., Parriaux, O.: Investigation of the polarization-dependent diffraction of deep dielectric rectangular transmission gratings illuminated in Littrow mounting. Appl. Opt. **46**, 819-826 (2007)
5. Dravins, D.: High-dispersion astronomical spectroscopy with holographic and ruled diffraction gratings. Appl. Opt. **17**, 404-414 (1978)
6. Harrison, G.R.: The Diffraction Grating—An Opinionated Appraisal. Appl. Opt. **12**, 2039-2049 (1973)
7. Habraken, S., Blanche, P.-A., Lemaire, P., Legros, N., Dekker, H., Monnet, G.: Volume Phase Holographic Gratings Made in Europe. Messenger **106**, 6-10 (2001)
8. Nguyen, H.T., Shore, B.W., Bryan, S.J., Britten, J. A., Boyd, R. D., Perry, M. D.: High-efficiency fused-silica transmission gratings. Opt. Lett. **22**, 142-144 (1997)
9. PTB Measurement Report.: Determination of the transmittance efficiency of an optical grating in the wavelength range of 300 to 410 nm. PTB-4.51-4093591 (2018). private communication
10. Shankoff, T.A.: Phase holograms in dichromated gelatine. Appl. Opt. **7**, 2101-2105 (1968)
11. Zanutta, A., Cristiani, S., Atkinson, D., Baldini, V., Balestra, A., Barbuy, B., Bawden Macanhan, V., Calcines, A., Case, S., Castilho, B., Cescutti, G., Cirami, R., Coretti, I., Covino, S., Cupani, G., Dekker, H., Di Marcantonio, P., D'Odorico, V., Ellis, S., Ernandes, H., Evans, C., Feger, T., Feiz, C., Franchini, M., Genoni, M., Gneiding, C., Kaluszynski, M., Landoni, M., Lawrence, J., Lunney, D., Miller, C., Mohanan, M., Molaverdikhani, K., O'Brien, K., Opitom, C., Pariani, G., Piranomonte, S., Quirrenbach, A., Redaelli, E., Riva, M., Robertson, D., Rothmaier, F., Seifert, W., Smiljanic, R., Stürmer, J., Stilz, I., Verducci, O., Waring, C., Watson, S., Wells, M., Xu, W., Zafar, T., Zheng J.: CUBES Phase-A design overview. Ex. Astr. (2022)

12. Zeitner, U.D., Oliva, M., Fuchs, F., Michaelis, D., Benkenstein, T., Harzendorf, T., Kley, E.-B.: High-performance diffraction gratings made by e-beam lithography. Appl. Phys. A **109**(4), 789-796 (2012)
13. Zeitner, U.D., Fuchs, F., Kley, E.-B.: High-performance dielectric diffraction gratings for space applications. Proc. SPIE 8450, Modern Technologies in Space- and Ground-based Telescopes and Instrumentation II, 84502Z (2012)
14. Zeitner, U.D., Fuchs, F., Kley, E.-B., Tünnermann, A.: High-refractive-index gratings for spectroscopic and laser applications. In: High Contrast Metastructures III, vol. 8995 (International Society for Optics and Photonics), p. 899504 (2014)